\documentclass[12pt]{amsart}

\usepackage{amssymb}

\theoremstyle{plain}

\numberwithin{equation}{section}

\begin{document}
\title{Epsilon coherent states with polyanalytic coefficients for the
harmonic oscillator}
\author{Zouha\"{\i}r Mouayn}
\maketitle
\address{{\footnotesize 
\begin{center}
Department of Mathematics, Faculty of Sciences and
Technics (M'Ghila)\\ BP. 523, B\'{e}ni Mellal, Morocco
\end{center}
}}

\begin{abstract}
We construct a \ new class of coherent states indexed by points $z$ of the
complex plane $\mathbb{\ }$and depending on two positive parameters $m$ and $%
\varepsilon >0$ by replacing the coefficients $z^{n}/\sqrt{n!}$ of the
canonical coherent states by polyanalytic functions$.$ These states solve
the identity of the states Hilbert space of the harmonic oscillator at the
limit $\varepsilon \rightarrow 0^{+}$ and obey a thermal stability property.
Their wavefunctions are obtained in a closed form and their associated
Bargmann-type transform is also discussed.
\end{abstract}

\section{Introduction}

In general, coherent states (CS) are a specific overcomplete family of
vectors in the Hilbert space of the problem that describes the quantum
phenomena and solves the identity of this Hilbert space. These states have
long been known for the harmonic oscillator (HO) potential and their
properties have frequently been taken as models for defining this notion for
other models \cite{AAG}-\cite{D}. The HO potential finds application in the
description of vibrational modes in nuclei, atoms, molecules and crystal
lattices.

Here, the CSs for HO potential we are introducing are quite different from
the existing ones in the above literature and are simply obtained by
adopting a general Hilbertian probabilistic scheme $\left[ 3\right] $
reminiscent to the classical construction of the Bargmann transform $\left[
22\right] $. Our procedure can be described as follows. In $\left[ 5\right] $
we have introduced a family of CS for the HO potential through
superpositions of the corresponding eigenstates where the role of
coefficients $z^{n}/\sqrt{n!}$ of the canonical CS was played by
coefficients 
\begin{equation}
\Phi _{n}^{m}\left( z\right) :=\left( -1\right) ^{m\wedge n}(m\wedge
n)!\left\vert z\right\vert ^{\left\vert m-n\right\vert }e^{-i(m-n)\arg
z}L_{m\wedge n}^{(\left\vert m-n\right\vert )}\left( z\overline{z}\right)
,z\in \mathbb{C},\text{ }n=0,1,...\text{,}  
\label{1.1}
\end{equation}%
where $L_{n}^{(\alpha )}(.)$ denotes the Laguerre polynomial $\left[ 14%
\right] $ and $m\wedge n=\min \left( m,n\right) .$ To be more precise, the $%
\left\{ \Phi _{n}^{m}\left( z\right) \right\} $ constitute an orthonormal
basis of a true polyanalytic space attached to a fixed $m$th Landau level 
\cite{AF}. Here, we proceed by modifying the coefficients $\left( \ref{1.1}\right) 
$ by a factor $e^{-n\varepsilon }$ . This defines what we call $epsilon$
coherent states and we denote by $\varepsilon $-CS for brevity. In fact
these $\varepsilon $-CS solve an $\varepsilon $-identity operator which has
the advantage of being a compact and trace class operator. The latter
becomes the identity operator of the Hilbert space $L^{2}\left( \mathbb{R}%
\right) $ at the limit $\varepsilon \rightarrow 0^{+}$. This can be proved
essentially by using a result on the Poisson kernel for Hermite polynomials,
which is due to Muckenhoopt $\left[ 11\right] $.

On the physical side, we can interpret the number $\varepsilon $ as the
usual parameter $\beta =1/k_{B}T$ of statistical physics where $k_{B}$ is
the Boltzmann constant and $T$ is the temperature. Therefore the resolution
of the $\varepsilon $-identity operator gives in fact (up to a normalization
factor) the thermodynamical quantum density for a HO potential with the form 
$\widehat{\rho }=$ $\sum_{n=0}^{\infty }n\left\vert n\right\rangle
\left\langle n\right\vert .$ These $\varepsilon $-CS also obey a thermal
stability property. Furthermore, the method we are using, which is similar
to the one used in \cite{Mo2}, makes possible to obtain a closed form for
the $\varepsilon $-CS allowing to define a Bargmann-type transform, say $%
B_{m}^{\varepsilon }$. The latter one can be considered as generalization
with the respect to parameter $\varepsilon $ of the true-polyanalytic $m$%
-Bargmann transform \cite{Va}-\cite{AF}.

The paper is organized as follows. In Section 2, we recall briefly some
needed fact on polyanalytic functions on the complex plane. Section 3 deals
with the coherent states formalism we will be using. \ In section 4 we
construct $\varepsilon $-coherent states and we show that they solve an $%
\varepsilon $-identity which becomes the identity of the states Hilbert
space at the limit $\varepsilon \rightarrow 0^{+}$. In Section 5, we give a
closed form for the constructed states an we discuss their associated
Bargmann-type transform.

\section{Polyanalytic functions on $\mathbb{C}$}

The Bargmann-Fock space $\mathbf{F}^{m+1}\left( \mathbb{C}\right) $ of 
\textit{polyanalytic} functions consists of all functions $F\left( z\right) $
satisfying the equation 
\begin{equation}
\left( \frac{\partial }{\partial \overline{z}}\right) ^{m+1}F\left( z\right)
=0 \label{2.1}
\end{equation}%
and such that 
\begin{equation}
\int_{\mathbb{C}}\left\vert F\left( z\right) \right\vert ^{2}e^{-z\overline{z%
}}d\mu \left( z\right) <+\infty 
\end{equation}%
where $d\mu \left( z\right) $ denotes the Lebesgue measure on $\mathbb{C}$.
Functions satisfying $\left( \ref{2.1}\right) $ are known as polyanalytic
functions of order $m+1.$ Since Eq.$\left( \ref{2.1}\right) $ generalizes the
Cauchy-Riemann equation 
\begin{equation}
\frac{\partial }{\partial \overline{z}}F\left( z\right) =0, 
\end{equation}%
then the space $\mathbf{F}^{m+1}\left( \mathbb{C}\right) $ is a
generalization of the well known Bargmann-Fock space $\mathcal{F}\left( 
\mathbb{C}\right) $ of entire Gaussian-square integrable functions on $%
\mathbb{C}.$ That is, for $m=0$, $\mathbf{F}^{1}\left( \mathbb{C}\right)
\equiv \mathcal{F}\left( \mathbb{C}\right) $, see $\left[ 12\right] $ , \cite%
{AF}. Polyanalytic functions inherit some of the properties of analytic
functions, in an nontrivial form. However, many of the properties break down
once we leave the analytic setting. For instance, while nonzero entire
functions do not have sets of zeros with an accumulation point, polyanalytic
functions can vanish along closed curves. To illustrate such situation, take 
$F\left( z\right) :=z\overline{z}-1$, a polyanalytic function of order 2.

Now, if we look at the so-called \textit{true} polyanalytic Fock spaces \cite%
{Va} ,\cite{A} which will be denoted here by ${\mathcal{A}}_{l}^{2}\left( 
\mathbb{C}\right) ,$ $l=0,1,...,m.$ These spaces are related to the
polyanalytic Fock space $\mathbf{F}^{m+1}\left( \mathbb{C}\right) $ by the
orthogonal decomposition (\cite{AF},\cite{Va},\cite{A}): 
\begin{equation}
\mathbf{F}^{m+1}\left( \mathbb{C}\right) ={\mathcal{A}}_{0}^{2}\left( 
\mathbb{C}\right) \oplus {\mathcal{A}}_{1}^{2}\left( \mathbb{C}\right)
\oplus ...\oplus {\mathcal{A}}_{m}^{2}\left( \mathbb{C}\right) . 
\end{equation}%
\ \ \ Moreover, for each fixed $m\in \mathbb{Z}_{+}$, the true polyanalytic
Bargmann-Fock space ${\mathcal{A}}_{m}^{2}\left( \mathbb{C}\right) $ admits
a nice realization as an eigenspace (\cite{AIM}): 
\begin{equation}
{\mathcal{A}}_{m}^{2}\left( \mathbb{C}\right) :=\left\{ f\in L^{2}\left( 
\mathbb{C},e^{-z\overline{z}}d\mu \right) ,~\widetilde{\Delta }f=mf\right\} \label{2.5}
\end{equation}%
of the second order differential operator 
\begin{equation}
\quad \widetilde{\Delta }:=-\frac{\partial ^{2}}{\partial z\partial 
\overline{z}}+\overline{z}\frac{\partial }{\partial \overline{z}}. 
\end{equation}
This operator constitutes, in suitable units and up to additive constant, a
realization in the Hilbert space $L^{2}\left( \mathbb{C},e^{-z\overline{z}%
}d\mu \right) $ of the Schr\"{o}dinger operator with uniform magnetic field
in $\mathbb{C}$. Its spectrum consists of eigenvalues $\lambda _{m}$ of
infinite multiplicity (\textit{Euclidean} \textit{Landau levels}) of the
form $\lambda _{m}:=m,$ $m\in \mathbb{Z}_{+}$. The space in $\left(
\ref{2.5}\right) $ admits an orthogonal basis whose elements are expressed by 
{\small 
\begin{equation}
\Phi _{n}^{m}\left( z\right) :=\left( -1\right) ^{m\wedge n}(m\wedge
n)!\left\vert z\right\vert ^{\left\vert m-n\right\vert }e^{-i(m-n)\arg
z}L_{m\wedge n}^{(\left\vert m-n\right\vert )}\left( z\overline{z}\right)
,z\in \mathbb{C},\text{ }n=0,1,...\text{,} \label{2.7}
\end{equation}%
}in terms of Laguerre polynomials $\left[ 19\right] :$ 
\begin{equation}
L_{j}^{\left( \alpha \right) }\left( t\right) =\sum\limits_{k=0}^{j}\frac{%
\Gamma \left( \alpha +j+1\right) }{\Gamma \left( \alpha +k+1\right) }\frac{%
\left( -t\right) ^{k}}{\left( j-k\right) !k!},\text{ \ }\alpha >-1  
\end{equation}%
with the orthogonality relations (with respect to the scalar product in $%
L^{2}\left( \mathbb{C},e^{-z\overline{z}}d\mu \right) $) given by 
\begin{equation}
\langle \Phi _{n}^{m}\left\vert \Phi _{j}^{m}\right\rangle =\pi m!n!\delta
_{n,j}=\pi m!n!\delta _{n,j}. \label{2.9}
\end{equation}%
where $\delta _{j,k}$ denotes the Kronecker symbol. Direct calculations
using $\left( \ref{2.7}\right) $,$\left( \ref{2.9}\right) $ together with a known
summation formula of the product of Laguerre polynomials allow to obtain the
reproducing kernel of the Hibert space ${{\mathcal{A}}}_{m}^{2}\left( 
\mathbb{C}\right) $ with the form (\cite{Mo3}) : 
\begin{equation}
K_{m}\left( z,w\right) =\pi ^{-1}e^{z\overline{w}}L_{m}^{\left( 0\right)
}\left( \left\vert z-w\right\vert ^{2}\right) \text{, \ \ \ \ }z,w\in 
\mathbb{C}. 
\end{equation}%
More information on these spaces and applications to signal analysis and
physics can be found in \cite{AF}, \cite{ABGM}, \cite{HH} and references
therein.

\section{Epsilon coherent states}

In this section, we will review a generalization of canonical CS by
considering a kind of the identity resolution that we obtain at the zero
limit with respect to a parameter $\varepsilon >0$. Their formalism can be
found in \cite{Mo2} where new families of CS attached to the Hamiltonian
with pseudo-harmonic oscillator potential were constructed.\smallskip 
\vspace{0in}

\textbf{Definition 3.1} . \textit{Let }$\mathcal{H}$\textit{\ be a (complex,
separable, infinite-dimensional) Hilbert space with an orthonormal basis }$%
\left\{ \psi _{n}\right\} _{n=0}^{\infty }.$\textit{\ Let }$\mathfrak{D}$%
\textit{\ }$\subseteq \mathbb{C}$\textit{\ be an open subset of }$\mathbb{C}$%
\textit{\ and let }$c_{n}:\mathfrak{D}\rightarrow \mathbb{C};n=0,1,2,\cdots
, $\textit{\ be a sequence of complex functions. Define } 
\begin{equation}
\left\vert z,\varepsilon \right\rangle :=\left( \mathcal{N}_{\varepsilon
}\left( z\right) \right) ^{-\frac{1}{2}}\sum\limits_{n=0}^{+\infty }\frac{%
\overline{c_{n}\left( z\right) }}{\sqrt{\sigma _{\varepsilon }\left(
n\right) }}\left\vert \psi _{n}\right\rangle \label{3.1}
\end{equation}%
\textit{where }$\mathcal{N}_{\varepsilon }\left( z\right) $\textit{\ is a
normalization factor and }$\sigma _{\varepsilon }\left( n\right) $\textit{; }%
$n=0,1,2,\cdots ,$\textit{\ a sequence of positive numbers depending on }$%
\varepsilon >0$\textit{. The vectors} $\left\{ \left\vert z,\varepsilon
\right\rangle ,z\in \mathfrak{D}\right\} $\textit{\ are said to form a set
of epsilon coherent states if\newline
}$\left( i\right) $\textit{\ for each fixed }$z\in $\textit{$\mathfrak{D}$ }%
\textit{\ and }$\varepsilon >0,$\textit{\ the state in ($\ref{3.1}$)} \textit{\ is
normalized, that is }$\left\langle z,\varepsilon |z,\varepsilon
\right\rangle _{\mathcal{H}}=1,$\textit{\newline
}$\left( ii\right) $ \textit{the following resolution of the identity
operator on }$\mathcal{H}$\textit{\ } 
\begin{equation}
\lim_{\varepsilon \rightarrow 0^{+}}\int_{\mathfrak{D}}\left\vert
z,\varepsilon \right\rangle \left\langle z,\varepsilon \right\vert d\mu
_{\varepsilon }\left( z\right) =\mathbf{1}_{\mathcal{H}} \label{3.2}
\end{equation}%
\textit{is satisfied with an appropriately chosen measure }$d\mu
_{\varepsilon }$.

\smallskip

\vspace{0in}In the above definition, the Dirac's\textit{\ bra-ket} notation $%
\left\vert z,\varepsilon \right\rangle \left\langle z,\varepsilon
\right\vert $ in $\left( \ref{3.2}\right) $ means the rank-one operator $\varphi
\mapsto \left\vert z,\varepsilon \right\rangle \left\langle z,\varepsilon
|\varphi \right\rangle _{\mathcal{H}}$, $\varphi \in \mathcal{H}.$ Also, the
limit in $\left( \ref{3.2}\right) $ is to be understood as follows. \ We define
the integral of rank-one operators as being the linear operator 
\begin{equation}
\mathcal{O}_{\varepsilon }\left[ \varphi \right] \left( \bullet \right)
:=\int_{\mathfrak{D}}\langle \bullet \left\vert z,\varepsilon \right\rangle
\langle z,\varepsilon |\varphi \rangle d\mu _{\varepsilon }\left( z\right).\label{3.3}
\end{equation}%
Then, the above limit is pointwise meaning $\mathcal{O}_{\varepsilon }\left[
\varphi \right] \left( \bullet \right) \rightarrow $ $\varphi \left( \bullet
\right) $ as $\varepsilon \rightarrow 0^{+}$, \textit{almost every where }%
with respect to $\left( \bullet \right) .$ Here, we should mention that the
usual way is to understand the integral 
\begin{equation}
\int_{\mathfrak{D}}\left\vert z,\varepsilon \right\rangle \langle
z,\varepsilon |d\mu _{\varepsilon }\left( z\right)  
\end{equation}%
in the weak sense, see for instance ($\left[ 5\right] $, p.8). Namely, it is
the sesquilinear form 
\begin{equation}
B_{\varepsilon }\left( \phi ,\psi \right) :=\int_{\mathfrak{D}}\langle \phi
\left\vert z,\varepsilon \right\rangle \langle z,\varepsilon |\psi \rangle
d\mu _{\varepsilon }\left( z\right) . \label{3.5}
\end{equation}%
Choosing this way, one has to check that the form $\left( \ref{3.5}\right) $ is
bounded so that the Riesz lemma ensures the existence of a unique bounded
operator $\mathcal{O}_{\varepsilon }$ satisfying $B_{\varepsilon }\left(
\phi ,\psi \right) =\langle \phi |\mathcal{O}_{\varepsilon }\left[ \psi %
\right] \rangle $. In our framework the resolution of the identity reads $%
\lim_{\varepsilon \rightarrow 0}$ $B_{\varepsilon }\left( \phi ,\psi \right)
=\langle \phi |\psi \rangle $ meaning that $\lim_{\varepsilon \rightarrow 0}%
\mathcal{O}_{\varepsilon }=\mathbf{1}_{\mathcal{H}}$ in the weak operator
topology.\newline

Note also that the above expression $\left( \ref{3.1}\right) $\ can be viewed as a
generalization of the series expansion of the canonical (anti-holomorphic)
coherent states 
\begin{equation}
\left\vert z\right\rangle :=\left( e^{z\bar{z}}\right) ^{-\frac{1}{2}%
}\sum_{n=0}^{+\infty }\frac{\overline{z}^{n}}{\sqrt{n!}}\left\vert \varphi
_{n}\right\rangle ;\quad z\in \mathbb{C},  
\end{equation}%
where $\left\{ \left\vert \varphi _{n}\right\rangle \right\} $ is an
orthonormal basis in $L^{2}\left( \mathbb{R}\right) $, which consists of
eigenstates of the Hamiltonian with the HO potential $-\partial
_{x}^{2}+x^{2}$ given by 
\begin{equation}
\varphi _{n}\left( x\right) =(\sqrt{\pi }2^{n}n!)^{-1/2}e^{-\frac{1}{2}%
x^{2}}H_{n}(x)  \label{3.7}
\end{equation}%
in terms of the Hermite polynomial 
\begin{equation}
H_{n}(x)=\sum_{k=0}^{\lfloor n/2\rfloor }\frac{\left( -1\right) ^{k}n!}{%
k!\left( n-2k\right) !}\left( 2x\right) ^{n-2k} 
\end{equation}%
see $\left[ 19\right] $. Here, the notation $\lfloor a\rfloor $ means the
greatest integer not exceeding $a$.
\newpage
\section{Epsilon CS with polyanalytic coefficients for the HO potential}

We now construct a class of $\varepsilon $-CS indexed by points $z\in 
\mathbb{C}$ and depending on two parameteres $m$ and $\varepsilon $ by
replacing the coefficients $z^{n}/\sqrt{n!}$ of the canonical coherent
states by polyanalytic coefficients as mentioned in the introduction.\newline

\textbf{Definition 4.1.} \textit{Define a set of states labeled by points }$%
z\in \mathbb{C}$\textit{\ and depending on two parameters }$m$\textit{\ and }%
$\varepsilon >0$\textit{\ by the following superposition} 
\begin{equation}
\left\vert z;m,\varepsilon \right\rangle :=\left( \mathcal{N}_{m,\varepsilon
}\left( z\right) \right) ^{-\frac{1}{2}}\sum\limits_{n=0}^{+\infty }\frac{%
\overline{\Phi _{n}^{m}\left( z\right) }}{\sqrt{\sigma _{\varepsilon
,m}\left( n\right) }}\left\vert \varphi _{n}\right\rangle  \label{4.1}
\end{equation}%
\textit{where}\emph{\ }$\mathcal{N}_{m,\varepsilon }\left( z\right) $\emph{\ 
}\textit{is a normalization factor, }$\sigma _{m,\varepsilon }(n)$\emph{\ }%
\textit{are positive numbers given by} 
\begin{equation}
\sigma _{m,\varepsilon }\left( n\right) :=\pi m!n!e^{n\varepsilon },\ \
n=0,1,2,\cdots ,  
\end{equation}%
\textit{and} $\left\{ \left\vert \varphi _{n}\right\rangle \right\} $ 
\textit{is the orthonormal basis of }$\mathcal{H}=L^{2}\left( \mathbb{R}%
\right) $,\textit{\ consisting of eigenstates of the harmonic oscillator as
given in (3.7)}.\newline
\vspace{0in}In the next result (see Appendix A) we give the overlap relation
between two $\varepsilon $-CS.

\textbf{Proposition 4.1.}\textit{\ Let }$m\in \mathbb{Z}_{+}$\textit{\ and }$%
\varepsilon >0$\textit{. Then, for every }$z$,$w$\textit{\ in }$\mathbb{C}$%
\textit{, the overlap relation between two }$\varepsilon $-\textit{CS is
expressed as}%
\begin{equation}
\langle z;m,\varepsilon \left\vert w;m,\varepsilon \right\rangle
_{L^{2}\left( \mathbb{R}\right) }=\frac{\exp \left( e^{-\varepsilon }z%
\overline{w}-m\varepsilon \right) }{\pi \sqrt{\mathcal{N}_{m,\varepsilon
}\left( z\right) \mathcal{N}_{m,\varepsilon }\left( w\right) }}L_{m}^{\left(
0\right) }\left( \left( ze^{-\varepsilon }-w\right) \left( \overline{z}%
e^{\varepsilon }-\overline{w}\right) \right) \label{4.3}
\end{equation}%
\textit{where the normalization factor is given by } 
\begin{equation}
\mathcal{N}_{m,\varepsilon }\left( z\right) =\pi ^{-1}\exp \left(
e^{-\varepsilon }z\overline{z}-m\varepsilon \right) L_{m}^{\left( 0\right)
}\left( 2\left( 1-\cosh \varepsilon \right) z\overline{z}\right)\label{4.4} 
\end{equation}%
\textit{in terms of the Laguerre polynomial }$L_{m}^{\left( 0\right) }\left(
.\right) $.\medskip

\textbf{Corollary 4.1. }\ \textit{At the limit }$\varepsilon \rightarrow
0^{+}$, \textit{the overlap relation (4.3) gives the normalized reproducing
kernel of the true polyanalytic Fock space }${{\mathcal{A}}}_{m}^{2}\left( 
\mathbb{C}\right) $\textit{. That is, } 
\begin{equation}
\lim_{\varepsilon \rightarrow 0^{+}}\langle z;m,\varepsilon \left\vert
w;m,\varepsilon \right\rangle _{L^{2}\left( \mathbb{R}\right) }=\frac{%
K_{m}\left( z,w\right) }{\sqrt{K_{m}\left( z,z\right) K_{m}\left( w,w\right) 
}} 
\end{equation}%
\textit{where} $K_{m}\left( z,w\right) $ \textit{is given explicitly by} 
\textit{(2.10).\medskip }

We now proceed to determine a measure of the form $\mathcal{N}%
_{m,\varepsilon }\left( z\right) d\eta \left( z\right) $ with respect to
which the $\varepsilon $-CS satisfy a resolution of an $\varepsilon $%
-identity operator and where $d\eta \left( z\right) $ is not $\varepsilon $%
-independent.

\textbf{Proposition 4.2. }\textit{The }$\varepsilon $-CS \textit{solve an }$%
\varepsilon $\textit{-identity operator as follows \vspace{0in}}%
\begin{equation}
 _{\mathbb{C}}\left\vert z;m,\varepsilon \right\rangle \langle
z;m,\varepsilon |d\mu _{m,\varepsilon }\left( z\right) =e^{-\varepsilon 
\mathbf{H}} 
\end{equation}%
\textit{where} $\mathbf{H=}\sum_{n=0}^{+\infty }n\left\vert \varphi
_{n}\right\rangle \langle \varphi _{n}|$ \textit{and} 
\begin{equation}
d\mu _{m,\varepsilon }\left( z\right) =e^{-z\overline{z}}\pi ^{-1}\exp
\left( e^{-\varepsilon }z\overline{z}-m\epsilon \right) L_{m}^{\left(
0\right) }\left( 2\left( 1-\cosh \varepsilon \right) z\overline{z}\right)
d\mu \left( z\right) 
\end{equation}%
\textit{with} $d\mu \left( z\right) $ \textit{being the Lebesgue measure on }%
$\mathbb{C}$.

\textbf{Proof.} Let us assume that the measure takes the form 
\begin{equation}
d\mu _{m,\varepsilon }\left( z\right) =\mathcal{N}_{m,\varepsilon }\left(
z\right) \rho \left( z\right) d\mu \left( z\right)  
\end{equation}%
where $\rho \left( z\right) $ is an auxiliary density to be determined. Let $%
\varphi \in L^{2}\left( \mathbb{R}\right) $ and let us start according to $%
\left( \ref{3.3}\right) $ by writing 
\begin{eqnarray}
\mathcal{O}_{m,\varepsilon }\left[ \varphi \right] &:=&\left( _{
\mathbb{C}}\left\vert z;m,\varepsilon \right\rangle \langle z;m,\varepsilon
|d\mu _{m,\varepsilon }\left( z\right) \right) \left[ \varphi \right] 
\\
&=& _{\mathbb{C}}\langle \varphi \left\vert z;m,\varepsilon
\right\rangle \langle z;m,\varepsilon |d\mu _{m,\varepsilon }\left( z\right)
\\
&=& \int\limits_{\mathbb{C}}\langle \varphi \mid \left( \mathcal{N}_{m,\varepsilon }\left( z\right) \right)^{-\frac{1}{2}}\sum\limits_{n=0}^{+\infty }\frac{\overline{\Phi _{n}^{m}\left( z\right) }}{\sqrt{\sigma_{\varepsilon ,m}\left( n\right) }}\left\vert \varphi_{n}\right\rangle
\rangle \langle z;m,\varepsilon |d\mu _{m,\varepsilon }\left( z\right) 
\\
&=& \int\limits_{\mathbb{C}}\sum\limits_{n=0}^{+\infty }\frac{\overline{\Phi
_{n}^{m}\left( z\right) }}{\sqrt{\sigma _{\varepsilon ,m}\left( n\right) }}%
\langle \varphi \left\vert \varphi _{n}\right\rangle \rangle \langle
z;m,\varepsilon |\left( \mathcal{N}_{m,\varepsilon }\left( z\right) \right)
^{-\frac{1}{2}}d\mu _{m,\varepsilon }\left( z\right)  \\
&=&\left( \sum\limits_{n,j=0}^{+\infty }\int\limits_{\mathbb{C}}\frac{%
\overline{\Phi _{n}^{m}\left( z\right) }\Phi _{j}^{m}\left( z\right) }{\sqrt{%
\sigma _{\varepsilon ,m}\left( n\right) }\sqrt{\sigma _{\varepsilon
,m}\left( j\right) }}\left\vert \varphi _{n}\right\rangle \langle \varphi
_{j}|\left( \mathcal{N}_{m,\varepsilon }\left( z\right) \right) ^{-1}d\mu
_{m,\varepsilon }\left( z\right) \right) \left[ \varphi \right] . \label{4.13}
\end{eqnarray}
Replace $d\mu _{m,\varepsilon }\left( z\right) =\mathcal{N}_{m,\varepsilon
}\left( z\right) \rho \left( z\right) d\mu \left( z\right) ,$ then Eq.$%
\left( \ref{4.13}\right) $ takes the form 
\begin{equation}
\mathcal{O}_{m,\varepsilon }=\sum\limits_{n,j=0}^{+\infty }e^{-\left(
n+j\right) \frac{\varepsilon }{2}}\left[ \int\limits_{\mathbb{C}}\frac{\Phi
_{j}^{m}\left( z\right) \overline{\Phi _{n}^{m}\left( z\right) }}{\sqrt{\pi
m!j!}\sqrt{\pi m!n!}}\rho \left( z\right) d\mu \left( z\right) \right]
\left\vert \varphi _{n}\right\rangle \langle \varphi _{j}|. \label{4.17}
\end{equation}%
We recall the orthogonality relations $\left( \ref{2.9}\right) $ $:$ 
\begin{equation}
\langle \Phi _{n}^{m}\left\vert \Phi _{j}^{m}\right\rangle =\pi m!n!\delta
_{n,j}. 
\end{equation}%
This suggests us to set 
\begin{equation}
\rho \left( z\right) :=e^{-z\overline{z}},z\in \mathbb{C}.  
\end{equation}%
Therefore, the operator in $\left( 4.14\right) $ takes the form \qquad\ 
\begin{equation}
\mathcal{O}_{m,\varepsilon }\left[ \varphi \right] \equiv \mathcal{O}%
_{\varepsilon }\left[ \varphi \right] =\sum\limits_{n=0}^{+\infty
}e^{-n\varepsilon }\left( \left\vert \varphi _{n}\right\rangle \langle
\varphi _{n}|\right) \left[ \varphi \right] . 
\end{equation}%
By defining a Hamiltonian operator of the harmonic oscillator type via the
discrete spectral resolution $\mathbf{H=}\sum_{n=0}^{+\infty }n\left\vert
\varphi _{n}\right\rangle \langle \varphi _{n}|,$ then Eq.$\left(
\ref{4.17}\right) $ also reads $\mathcal{O}_{\varepsilon }\left[ \varphi \right]
=e^{-\varepsilon \mathbf{H}}\left[ \varphi \right] ,$ $\varphi \in \mathcal{H%
}$. $\square $

In the next result (see Appendix B) we state the resolution of the identity
operator.

\textbf{Proposition 4.3}. \textit{The }$\varepsilon $-CS \textit{satisfy the
following resolution of the identity } 
\begin{equation}
\lim_{\varepsilon \rightarrow 0^{+}} \,_{\mathbb{C}}| z;m,\varepsilon \rangle  \langle z;m,\varepsilon |d\mu _{m,\varepsilon}\left( z\right) =\mathbf{1}_{L^{2}\left( \mathbb{R}\right) }  
\end{equation}
\textit{where } $d\mu _{m,\varepsilon }\left( z\right) $\textit{\ is the
measure given by} $\left( 4.7\right) .$

We close this section by mentionning the following property of these $%
\varepsilon -$CS.

\textbf{Proposition 4.4. }\textit{The }$\varepsilon $-CS\textit{\ obey the
following thermal stability property} 
\begin{equation}
e^{-\frac{1}{2}t\left( -\partial _{x}^{2}+x^{2}-\frac{1}{2}\right)
}\left\vert z;m,\varepsilon \right\rangle =\left( \frac{\mathcal{N}
_{m,\varepsilon +t}\left( z\right) }{\mathcal{N}_{m,\varepsilon }\left(
z\right) }\right) ^{\frac{1}{2}}\left\vert z;m,\varepsilon +t\right\rangle , \qquad t>0.
\end{equation}
\smallskip
\textbf{Proof. }On one hand, we write the spectral resolution of the heat
opertor $e^{-\frac{1}{2}t\widetilde{L}}$ associated with the shifted
harmonic oscillator $\widetilde{L}:$=$-\partial _{x}^{2}+x^{2}-\frac{1}{2}$
as 
\begin{equation}
e^{-\frac{1}{2}t\widetilde{L}}=\sum_{j=0}^{+\infty }e^{-\frac{1}{2}%
jt}\left\vert \varphi _{j}\right\rangle \langle \varphi _{j}|.  
\end{equation}%
On an other hand, we rewrite the $\varepsilon -$CS in $\left( \ref{4.1}\right) $
as 
\begin{equation}
\left\vert z;m,\varepsilon \right\rangle :=\left( \mathcal{N}_{m,\varepsilon
}\left( z\right) \right) ^{-\frac{1}{2}}\sum\limits_{n=0}^{+\infty }\gamma
_{n}^{\left( m\right) }\left( z\right) e^{-\frac{1}{2}n\varepsilon
}\left\vert \varphi _{n}\right\rangle ,  \label{4.21}
\end{equation}%
where $\gamma _{n}^{\left( m\right) }\left( z\right) :=\left( \pi
m!n!\right) ^{-\frac{1}{2}}\overline{\Phi _{n}^{m}\left( z\right) }.$ So
that writting the action of the heat operator $e^{-\frac{1}{2}t\widetilde{L}%
} $ on the form $\left( \ref{4.21}\right) ,$ we get successively 
\begin{eqnarray}
e^{-\frac{1}{2}t\widetilde{L}}\left\vert z;m,\varepsilon \right\rangle
&=&e^{-\frac{1}{2}t\widetilde{L}}\left( \left( \mathcal{N}_{m,\varepsilon }\left(
z\right) \right) ^{-\frac{1}{2}}\sum\limits_{n=0}^{+\infty }\gamma
_{n}^{\left( m\right) }\left( z\right) e^{-\frac{1}{2}n\varepsilon
}\left\vert \varphi _{n}\right\rangle \right)  \\
&=&\sum_{j=0}^{+\infty }e^{-\frac{1}{2}tj}\left\vert \varphi
_{j}\right\rangle \langle \varphi_{j}|\left( \left( \mathcal{N}%
_{m,\varepsilon }\left( z\right) \right) ^{-\frac{1}{2}}\sum\limits_{n=0}^{+%
\infty }\gamma _{n}^{\left( m\right) }\left( z\right) e^{-\frac{1}{2}%
n\varepsilon }\left\vert \varphi _{n}\right\rangle \right)   \\
&=&\left( \mathcal{N}_{m,\varepsilon }\left( z\right) \right) ^{-\frac{1}{2}%
}\sum_{j,n=0}^{+\infty }\gamma _{n}^{\left( m\right) }\left( z\right) e^{-%
\frac{1}{2}jt-\frac{1}{2}n\varepsilon }\left\vert \varphi _{j}\right\rangle
\langle \varphi _{j}|\varphi _{n}\rangle . 
\end{eqnarray}%
Using the orthonormality relation $\langle \varphi _{j}|\varphi _{k}\rangle
=\delta _{k,j}$, this action reduces to 
\begin{equation}
e^{-\frac{1}{2}t\widetilde{L}}\left\vert z;m,\varepsilon \right\rangle
=\left( \mathcal{N}_{m,\varepsilon }\left( z\right) \right) ^{-\frac{1}{2}%
}\sum_{n=0}^{+\infty }\gamma _{n}^{\left( m\right) }\left( z\right) e^{-%
\frac{1}{2}n\left( t+\varepsilon \right) }\left\vert \varphi
_{n}\right\rangle  
\end{equation}%
and it can also be written as 
\begin{equation}
e^{-\frac{1}{2}t\widetilde{L}}\left\vert z;m,\varepsilon \right\rangle
=\left( \frac{\mathcal{N}_{m,\varepsilon +t}\left( z\right) }{\mathcal{N}%
_{m,\varepsilon }\left( z\right) }\right) ^{\frac{1}{2}}\left\vert
z;m,\varepsilon +t\right\rangle\label{4.26}  
\end{equation}%
which means that, up to a factor depending on the labelling point $z,$ the
action of the heat operator $\exp \left( -\frac{1}{2}t\widetilde{L}\right) $
reproduces a similar state $\varepsilon $-CS, where $\varepsilon $ is
shifted by $t$. $\square \medskip $

\textbf{Remark. 4.1. \ }Eq.$\left( \ref{4.26}\right) $ means that these $%
\varepsilon $-CS satisfy a thermal stability (with respect to $\widetilde{L}%
\geq 0$). What also make this property possible is the linearity (with
respect to the integer index) of the spectrum of the Hamiltonian with HO
potential. \ A similar fact can be found in \cite{GK} where Gazeau and
Klauder introduced a real two parameters set of coherents, say $\left\{
\left\vert J,\gamma \right\rangle ,J\geq 0,\gamma \in \mathbb{R}\right\} $
associated with the discrete dynamics of a positive Hamiltonian $\widehat{H}$%
. One of the requirements for the their CS was the so-called temporal
stability meaning that $e^{-it\widehat{H}}\left\vert J,\gamma \right\rangle
=\left\vert J,\gamma +\omega t\right\rangle ,\omega $ can be taken equal to
one.
\newpage
\section{ A closed form for the $\protect\varepsilon $-CS}

In this section we will establish a closed form for the constructed $%
\varepsilon -$CS and we will discuss the associated Bargmann-type integral
transform.\smallskip

\textbf{Proposition 5.1.} \textit{Let} $m\in \mathbb{Z}_{+}$\textit{\ and }$%
\varepsilon >0$\textit{\ be fixed parameters. Then, the wavefunctions of the 
}$\varepsilon $-CS\textit{\ defined in }$\left( \ref{4.1}\right) $\textit{\ can be
written in a closed form as} 
\begin{equation}
\langle x\left\vert z;m,\varepsilon \right\rangle =\frac{\left( -1\right)
^{m}\left( e^{-\frac{1}{2}\varepsilon }/\sqrt{2} \right)^{m}}{\left( \sqrt{%
\pi }\right) ^{\frac{3}{2}}\sqrt{m!}}\frac{\exp \left( -\frac{1}{2}x^{2}+%
\sqrt{2}x\overline{z}e^{-\frac{1}{2}\varepsilon }-\frac{1}{2}e^{-\varepsilon
}\overline{z}^{2}\right)H_{m}\left( x-\frac{1}{%
\sqrt{2}}\left( e^{\frac{1}{2}\varepsilon }z+e^{-\frac{1}{2}\varepsilon }%
\overline{z}\right) \right) }{\sqrt{\pi ^{-1}\exp \left( e^{-\varepsilon }z%
\overline{z}-m\varepsilon \right) L_{m}^{\left( 0\right) }\left( 2\left(
1-\cosh \varepsilon \right) z\overline{z}\right) }} \label{5.1}  
\end{equation}
\textit{for every} $x\in \mathbb{R}$.\smallskip

\textbf{Proof. }We start by writing the expression of the wave function of $%
\varepsilon $-CS according to Definition $\left( \ref{4.1}\right) $ as 
\begin{equation}
\langle x\left\vert z;m,\varepsilon \right\rangle =\left( \mathcal{N}%
_{m,\varepsilon }\left( z\right) \right) ^{-\frac{1}{2}}\sum\limits_{n=0}^{+%
\infty }\frac{\overline{\Phi _{n}^{m}\left( z\right) }}{\sqrt{\sigma
_{\varepsilon ,m}\left( n\right) }}\varphi _{n}\left( x\right) ;\quad x\in 
\mathbb{R}. 
\end{equation}%
We have thus to look for a closed form of the series 
\begin{equation}
\mathcal{S}_{z}^{m,\varepsilon }\left( x\right) :=\sum\limits_{n=0}^{+\infty
}\frac{\overline{\Phi _{n}^{m}\left( z\right) }}{\sqrt{\sigma _{\varepsilon
,m}\left( n\right) }}\varphi _{n}\left( x\right) . \label{5.3} 
\end{equation}%
To do this, we start by replacing the coefficients $\Phi _{n}^{m}\left(
z\right) $ by their expression in $\left( \ref{2.7}\right) .$ So that Eq.$\left(
\ref{5.3}\right) $ reads 
\begin{equation}
\mathcal{S}_{z}^{m,\varepsilon }\left( x\right) =\sum\limits_{n=0}^{+\infty }%
\frac{e^{-\frac{1}{2}n\varepsilon }}{\sqrt{\pi n!m!}}\left( -1\right)
^{n\wedge m}(m\wedge n)!\left\vert z\right\vert ^{\left\vert m-n\right\vert
}e^{-i(m-n)\arg z}L_{m\wedge n}^{(\left\vert m-n\right\vert )}\left( z
\overline{z}\right) \varphi _{n}\left( x\right) ,  \label{5.4}
\end{equation}
where $m\wedge n:=$ $\min (m,n).$ Next, with the help of the identity $%
\left( A.7\right) $ on Laguerre polynomials, we are able to rewrite $\left(
\ref{5.4}\right) $ in the following form 
\begin{equation}
\mathcal{S}_{z}^{m,\varepsilon }\left( x\right) =\frac{\left( -1\right) ^{m}
\sqrt{m!}}{\sqrt{\pi }}\sum\limits_{n=0}^{+\infty }\frac{e^{-\frac{1}{2}
n\varepsilon }}{\sqrt{n!}}\overline{z}^{n-m}L_{m}^{(n-m)}\left( z\overline{z}\right)
\varphi _{n}\left( x\right) . \label{5.5}
\end{equation}%
Making use of the explicite expression $\left( \ref{3.7}\right) $ of the
eigenstates $\varphi _{n}\left( x\right) $, then the sum in $\left(
\ref{5.5}\right) $ becomes 
\begin{equation}
\mathcal{S}_{z}^{m,\varepsilon }\left( x\right) =\frac{\left( -1\right) ^{m}
\sqrt{m!}}{\left( \sqrt{\pi }\right) ^{\frac{3}{2}}\overline{z}^{m}}e^{-\frac{1}{2}
x^{2}}\sum\limits_{n=0}^{+\infty }\frac{\left( \left( 2e^{\varepsilon
}\right) ^{-\frac{1}{2}}\overline{z}\right) ^{n}}{n!}L_{m}^{(n-m)}\left( z\overline{z}
\right) H_{n}\left( x\right) . \label{5.6}
\end{equation}%
We now introduce the notation $\tau :=\left( 2e^{\varepsilon }\right) ^{-%
\frac{1}{2}}\overline{z}$ in $\left( \ref{5.6}\right) $ and we will be dealing with the sum 
\begin{equation}
G_{z}^{m,\varepsilon }\left( x\right) :=\sum\limits_{n=0}^{+\infty }\frac{
\tau ^{n}}{n!}H_{n}\left( x\right) L_{m}^{(n-m)}\left( z\overline{z}\right) .\label{5.7}
\end{equation}
Using the integral representation of Hermite polynomials (\cite{GR}, p.365): 
\begin{equation}
H_{p}\left( x\right) =\frac{e^{x^{2}}}{\sqrt{\pi }}\int\limits_{\mathbb{R}
}\left( 2iu\right) ^{p}e^{-2iux}e^{-u^{2}}du, 
\end{equation}
then the sum $\left( \ref{5.7}\right) $ may be written as 
\begin{equation}
G_{z}^{m,\varepsilon }\left( x\right) =\sum_{n=0}^{+\infty }\frac{1}{n!}
\left( \frac{e^{x^{2}}}{\sqrt{\pi }}\int\limits_{\mathbb{R}}\left( 2iu\tau
\right) ^{n}e^{-2iux}e^{-u^{2}}du\right) L_{m}^{(n-m)}\left( z\overline{z}
\right) \label{5.9}
\end{equation}
\begin{equation}
=\frac{e^{x^{2}}}{\sqrt{\pi }}\int\limits_{\mathbb{R}}e^{-2iux}e^{-u^{2}}
\left( \sum\limits_{n=0}^{+\infty }\left( 2iu\tau \right) ^{n}\frac{1}{n!}
L_{m}^{\left( n-m\right) }\left( z\overline{z}\right) \right) du. \label{5.10}
\end{equation}
The sum in $\left( \ref{5.10}\right)$ can also be presented as follows 
\begin{equation}
\sum_{n=0}^{+\infty }\left( \left( \frac{1}{\overline{z}}2iu\left(
2e^{\varepsilon }\right) ^{-\frac{1}{2}}\right) z\overline{z}\right) ^{n}
\frac{1}{n!}L_{m}^{\left( n-m\right) }\left( z\overline{z}\right) . \label{5.11}
\end{equation}
Making appeal to the following formula due to Deruyts (\cite{Bu}, p.142)$:$
\begin{equation}
\sum_{n=0}^{+\infty }\left( s\alpha \right) ^{n}\frac{1}{n!}L_{n}^{\left(
n-m\right) }\left( s\right) =\frac{s^{m}}{m!}\left( \alpha -1\right)
^{m}e^{s\alpha }
\end{equation}
for the parameters $\alpha =\frac{1}{\overline{z}}2iu\left( 2e^{\varepsilon
}\right) ^{-\frac{1}{2}}$ and $s=z\overline{z}$ then the sum $\left(
\ref{5.11}\right) $ reduces to 
\begin{equation}
\frac{z^{m}}{m!}\left( iu\sqrt{2}e^{-\frac{1}{2}\varepsilon }-z
\right) ^{m}\exp \left( \sqrt{2}iue^{-\frac{1}{2}\varepsilon }\overline{z}\right) . \label{5.13}
\end{equation}
Returning back to $\left( \ref{5.10}\right) $ and inserting the quantity $\left(
\ref{5.13}\right) $ then $\left( \ref{5.9}\right) $ reads 
\begin{equation}
G_{z}^{m,\varepsilon }\left( x\right) =\frac{\overline{z}^{m}e^{x^{2}}}{m!\sqrt{\pi }}
\int\limits_{\mathbb{R}}e^{-2iux}e^{-u^{2}}\left( iu\sqrt{2}e^{-\frac{1}{2}
\varepsilon }-z\right) ^{m}\exp \left( \sqrt{2}iue^{-\frac{1}{2}
\varepsilon }\overline{z}\right) du. \label{5.14}
\end{equation}
We now use the binomial formula 
\begin{equation}
\left( iu\sqrt{2}e^{-\frac{1}{2}\varepsilon }-z\right)
^{m}=\sum_{l=0}^{m}\left( 
\begin{array}{c}
m \\ 
l
\end{array}
\right) \left( iu\sqrt{2}e^{-\frac{1}{2}\varepsilon }\right) ^{l}\left( -
z\right) ^{m-l} 
\end{equation}
and we replace it in the right hand side of $\left( \ref{5.14}\right) $. We obtain
that 
\begin{equation}
G_{z}^{m,\varepsilon }\left( x\right) =\frac{\overline{z}^{m}e^{x^{2}}}{m!\sqrt{\pi }}%
\int\limits_{\mathbb{R}}e^{-2iux-u^{2}}\exp \left( \sqrt{2}iue^{-%
\frac{1}{2}\varepsilon }\overline{z}\right) \left( \sum_{l=0}^{m}\left( 
\begin{array}{c}
m \\ 
l
\end{array}
\right) \left( iu\sqrt{2}e^{-\frac{1}{2}\varepsilon }\right) ^{l}\left( -
z\right) ^{m-l}\right) du.  
\end{equation}%
This can also be written as 
\begin{equation}
G_{z}^{m,\varepsilon }\left( x\right) =\frac{\overline{z}^{m}e^{x^{2}}}{m!\sqrt{\pi }}%
\sum_{l=0}^{m}\left( 
\begin{array}{c}
m \\ 
l%
\end{array}
\right) \left( -z\right) ^{m-l}\int\limits_{\mathbb{R}
}e^{-2iux-u^{2}}\exp \left( \sqrt{2}iue^{-\frac{1}{2}\varepsilon }\overline{z}\right)
\left( iu\sqrt{2}e^{-\frac{1}{2}\varepsilon }\right) ^{l}du.  \label{5.17}
\end{equation}%
Now, the integral in $\left( \ref{5.17}\right) :$%
\begin{equation}
I_{l}=\left( \frac{e^{-\frac{1}{2}\varepsilon }}{\sqrt{2}}
\right)^l
\int\limits_{\mathbb{R}}\exp \left( -2iu\left( x-\overline{z}\frac{e^{-\frac{1}{2}
\varepsilon }}{\sqrt{2}}\right) \right) e^{-u^{2}}\left( i2u\right) ^{l}du 
\end{equation}
can be written by taking into account $\left( 5.8\right) $ as 
\begin{equation}
I_{l}=\left( \frac{e^{-\frac{1}{2}\varepsilon }}{\sqrt{2}}
\right)^l \sqrt{\pi }
\exp \left( -\left( x-\overline{z}\frac{e^{-\frac{1}{2}\varepsilon }}{\sqrt{2}}\right)
^{2}\right) H_{l}\left( x-\overline{z}\frac{e^{-\frac{1}{2}\varepsilon }}{\sqrt{2}}
\right) . \label{5.19}
\end{equation}
Replacing $\left( \ref{5.19}\right) $ in $\left( \ref{5.17}\right) $, we arrive at 
\begin{equation}
G_{z}^{m,\varepsilon }\left( x\right) =\frac{\overline{z}^{m}e^{x^2}}{m!}\left(\frac{e^{-\frac{1}{2}\varepsilon}}{\sqrt{2}}\right)^m\exp \left( -\left( x-\overline{z}\frac{e^{-\frac{1}{2}%
\varepsilon }}{\sqrt{2}}\right) ^{2}\right) \sum_{l=0}^{m}\left( 
\begin{array}{c}
m \\ 
l
\end{array}
\right) \left( -\sqrt{2}ze^{\frac{1}{2}\varepsilon}\right) ^{m-l}H_{l}\left( x-\overline{z}\frac{e^{-\frac{1}{2
}\varepsilon }}{\sqrt{2}}\right) .\label{5.20}  
\end{equation}%
Next, we apply the following identity to the last sum in $\left( \ref{5.20}\right) 
$ (\cite{MOS}, p.255): 
\begin{equation}
\sum_{l=0}^{m}\left( 
\begin{array}{c}
m \\ 
l
\end{array}
\right) \left( -2a\right) ^{m-l}H_{l}\left( t\right) =H_{m}\left( t-a\right)
. 
\end{equation}
Finally, summarizing the above calculations, we arrive at the announced
expression for the $\varepsilon -$CS in $\left( \ref{5.1}\right) $. $\square $

\section{The transform $\mathcal{B}_{m}^{\protect\varepsilon }$}

Naturally, once we have obtained a closed \ form for the $\varepsilon $-CS
we can look for the associated coherent states transform, say $%
B_{m}^{\varepsilon }$. In view of the definition $\left( \ref{4.1}\right) $, this
transform should map the space $L^{2}\left( \mathbb{R}\right) $ spanned by
eigenstates $\left\vert \varphi _{n}\right\rangle $ of the Hamiltonian with
the HO potential onto the $\varepsilon $\textit{-true polyanalytic} space ${{%
\mathcal{A}}}_{m}^{2,\varepsilon }\left( \mathbb{C}\right) $ which can be
defined as the subspace of \ $L^{2}\left( \mathbb{C},\pi^{-1}e^{-z\overline{z}}d\mu
\right) $ obtained as the closure of vector space spanned by all linear
combinations of the polyanalytic functions $z\mapsto \left( \sigma
_{\varepsilon ,m}\left( n\right) \right) ^{-\frac{1}{2}}\Phi _{n}^{m}\left(
z\right) $ with the normalized reproducing kernel

\begin{equation}
K_{m,\varepsilon }\left( z,w\right) :=\frac{\exp \left( e^{-\varepsilon }z
\overline{w}-m\varepsilon \right) }{ \sqrt{\mathcal{N}_{m,\varepsilon
}\left( z\right) \mathcal{N}_{m,\varepsilon }\left( w\right) }}L_{m}^{\left(
0\right) }\left( \left( ze^{-\varepsilon }-w\right) \left( \overline{z}
e^{\varepsilon }-\overline{w}\right) \right).
\end{equation}
Here, $\sigma_{\varepsilon ,m}(n)=m!n!e^{n\varepsilon}$ and $\mathcal{N}_{m,\varepsilon }\left( z\right) =\exp \left(
e^{-\varepsilon }z\overline{z}-m\varepsilon \right) L_{m}^{\left( 0\right)
}\left( 2\left( 1-\cosh \varepsilon \right) z\overline{z}\right)$.\\
\\
\textbf{Definition 6.1.} \textit{The Bargmann-type integral transform }$%
B_{m}^{\varepsilon }:L^{2}\left( \mathbb{R}\right) \rightarrow {{\mathcal{A}}%
}_{m}^{2,\varepsilon }\left( \mathbb{C}\right) $\textit{\ associated with
the }$\varepsilon $\textit{-CS is defined by} 
\begin{equation}
B_{m}^{\varepsilon }\left[ \varphi \right] \left( z\right) =\left( \mathcal{N
}_{m,\varepsilon }\left( z\right) \right) ^{\frac{1}{2}}\langle \varphi
\left\vert z;m,\varepsilon \right\rangle _{L^{2}\left( \mathbb{R}\right) } 
\end{equation}
\textit{and} \textit{explicitly as} 
\begin{equation*}
B_{m}^{\varepsilon }\left[ \varphi \right] \left( z\right) :=\frac{\left(
-1\right)^{m} e^{-\frac{1}{2}m\varepsilon } }{ 2^{m/2}\sqrt{m!}\pi^{1/4}}\int\limits_{%
\mathbb{R}}\exp \left( -\frac{x^2}{2}+\sqrt{2}xze^{-\frac{1}{2}\varepsilon} -\frac{e^{-\varepsilon}z^2}{2}\right) H_{m}\left(
x- \frac{e^{-\frac{1}{2}\varepsilon } \overline{z}}{\sqrt{2}}-\frac{e^{
\frac{1}{2}\varepsilon }z}{\sqrt{2}} \right) \varphi \left( x\right) dx
\end{equation*}%
\textit{for every} $z\in \mathbb{C}$.\smallskip

\smallskip At the limit $\varepsilon \rightarrow 0^{+},$ we recover the
coherent states transform $\left[ 5\right] :$ 
\begin{equation}
B_{m}^{0}:L^{2}\left( \mathbb{R}\right) \rightarrow A_{m}^{2}\left( \mathbb{C
}\right)  
\end{equation}
defined by\begin{equation*}
B_{m}^{0 }\left[ \varphi \right] \left( z\right) :=\frac{\left(
-1\right) ^{m} }{ 2^{m/2}\sqrt{m!}\pi^{1/4}}\int\limits_{%
\mathbb{R}}\exp \left( -\frac{x^2}{2}+\sqrt{2}xz -\frac{z^2}{2}\right)  \varphi \left( x\right) dx
\end{equation*}%
The latter one can also be written \cite{AF} as 
\begin{equation}
B_{m}^{0}\left[ \varphi \right] \left( z\right) \propto e^{\pi z\overline{z}
}\left( \partial _{z}\right) ^{m-1}\left( e^{-\pi z\overline{z}}B\left[
\varphi \right] \left( z\right) \right) 
\end{equation}
in terms of the transform 
\begin{equation}
B\equiv B_{0}^{0}:L^{2}\left( \mathbb{R}\right) \rightarrow A_{0}^{2}\left( 
\mathbb{C}\right) \equiv \mathcal{F}\left( \mathbb{C}\right)  
\end{equation}
\begin{equation*}
\varphi \mapsto B\left[ \varphi \right] \left( z\right) =\pi^{-1/4}\int\limits_{%
\mathbb{R}}\exp \left( -\frac{x^2}{2}+\sqrt{2}xz -\frac{z^2}{2}\right)  \varphi \left( x\right) dx
\end{equation*}%
which is the well known Bargmann transform \cite{Ba} .
\\
\\
\begin{center}
\textbf{Appendix A}
\end{center}

\textbf{Proof. }Using the orthogonality relations of the basis elements $
\left\{ \varphi _{n}\left( x\right) \right\} $ in $(\ref{3.7})$ the scalar product
in $L^{2}\left( \mathbb{R}\right) $ beteween two $\varepsilon $-CS can
written as 
\begin{equation}
\langle z;m,\varepsilon \left\vert w;m,\varepsilon \right\rangle
_{L^{2}\left( \mathbb{R}\right) }=\frac{Q_{\varepsilon }\left( z,w\right) }{%
\pi m!\sqrt{\mathcal{N}_{m,\varepsilon }\left( z\right) \mathcal{N}%
_{m,\varepsilon }\left( w\right) }}  \tag{A1}
\end{equation}%
where 
\begin{equation}
Q_{\varepsilon }\left( z,w\right) =\sum\limits_{n=0}^{+\infty }\frac{%
e^{-n\varepsilon }}{n!}\Phi _{n}^{m}\left( z\right) \overline{\Phi
_{n}^{m}\left( w\right) }.  \tag{A2}
\end{equation}%
Recalling the explicite expression $\left( \ref{2.7}\right) $ of the of the
polyanalytic coefficients$,$ we can split the sum in $\left( A2\right) $
into two part as 
\begin{equation*}
Q_{\varepsilon }\left( z,w\right) =\sum\limits_{n=0}^{m-1}e^{-n\varepsilon
}n!\left( \left\vert z\right\vert \left\vert w\right\vert \right) ^{\left(
m-n\right) }L_{n}^{\left( m-n\right) }\left( z\overline{z}\right)
L_{n}^{\left( m-n\right) }\left( w\overline{w}\right) e^{-i(m-n)\arg
z}e^{i(m-n)\arg w}
\end{equation*}%
\begin{equation}
+\sum\limits_{n=m}^{+\infty }\frac{e^{-n\varepsilon }}{n!}\left( m!\right)
^{2}\left( \left\vert z\right\vert \left\vert w\right\vert \right) ^{\left(
n-m\right) }L_{m}^{\left( n-m\right) }\left( z\overline{z}\right)
L_{m}^{\left( n-m\right) }\left( w\overline{w}\right) e^{-i(m-n)\arg
z}e^{i(m-n)\arg w}.  \tag{A3}
\end{equation}%
This quantity can also be decomposed as 
\begin{equation}
Q_{\varepsilon }\left( z,w\right) =Q_{\varepsilon }^{\left( <\infty \right)
}\left( z,w\right) +Q_{\varepsilon }^{\left( \infty \right) }\left(
z,w\right)  \tag{A4}
\end{equation}%
with a finite sum 
\begin{equation}
Q_{\varepsilon }^{\left( <\infty \right) }\left( z,w\right)
:=\sum\limits_{n=0}^{m-1}e^{-n\varepsilon }n!\left( \overline{z}w\right)
^{m-n}L_{n}^{\left( m-n\right) }\left( z\overline{z}\right) L_{n}^{\left(
m-n\right) }\left( w\overline{w}\right)  \tag{A5}
\end{equation}%
\begin{equation*}
-\sum\limits_{n=0}^{m-1}\frac{e^{-n\varepsilon }}{n!}\left( m!\right)
^{2}\left( z\overline{w}\right) ^{n-m}L_{m}^{\left( n-m\right) }\left( z%
\overline{z}\right) L_{m}^{\left( n-m\right) }\left( w\overline{w}\right)
\end{equation*}%
and an infinite sum 
\begin{equation}
Q_{\varepsilon }^{\left( \infty \right) }\left( z,w\right)
:=\sum\limits_{n=0}^{+\infty }\frac{e^{-n\varepsilon }}{n!}\left( m!\right)
^{2}\left( z\overline{w}\right) ^{n-m}L_{m}^{\left( n-m\right) }\left( z%
\overline{z}\right) L_{m}^{\left( n-m\right) }\left( w\overline{w}\right) . 
\tag{A6}
\end{equation}%
Making appeal to the identity ($\left[ S\right] $, p.98): 
\begin{equation}
L_{m}^{\left( -k\right) }\left( t\right) =\left( -t\right) ^{k}\frac{\left(
m-k\right) !}{m!}L_{m-k}^{\left( k\right) }\left( t\right) ,1\leq k\leq m 
\tag{A7}
\end{equation}%
for $k=j-m$ and $t=z\overline{z}$, we can check that the finite sum $%
Q_{\varepsilon }^{\left( <\infty \right) }\left( z,w\right) =0$. For the
infinite sum in $\left( A6\right) ,$ we rewrite it as 
\begin{equation}
Q_{\varepsilon }^{\left( \infty \right) }\left( z,w\right) =\frac{\left(
m!\right) ^{2}}{\left( z\overline{w}\right) ^{m}}\sum\limits_{n=0}^{+\infty }%
\frac{1}{n!}\left( z\overline{w}e^{-\varepsilon }\right) ^{n}L_{m}^{\left(
n-m\right) }\left( z\overline{z}\right) L_{m}^{\left( n-m\right) }\left( w%
\overline{w}\right) .  \tag{A8}
\end{equation}%
We now apply the Wicksell-Campbell-Meixner formula ($\left[ SM\right] $,
p.279)$:$%
\begin{equation}
\sum\limits_{n=0}^{+\infty }\frac{\zeta ^{n}}{n!}L_{l}^{\left( n-l\right)
}\left( X\right) L_{m}^{\left( n-m\right) }\left( Y\right) =e^{\zeta }\left(
\zeta -Y\right) ^{m-l}\frac{\zeta ^{l}}{m!}L_{l}^{\left( m-l\right) }\left(
-\left( X-\zeta \right) \left( Y-\zeta \right) \zeta ^{-1}\right)  \tag{A9}
\end{equation}%
with the notations $\zeta =e^{-\varepsilon }z\overline{w}$, $X=z\overline{z}%
,Y=w\overline{w}$ and $l=m.$ With this, Eq.$\left( A8\right) $ reduces to 
\begin{equation}
Q_{\varepsilon }^{\left( \infty \right) }\left( z,w\right)
=m!e^{-m\varepsilon }\exp \left( e^{-\varepsilon }z\overline{z}\right)
L_{m}^{\left( 0\right) }\left( \left( we^{\varepsilon }-z\right) \left( 
\overline{w}e^{-\varepsilon }-\overline{z}\right) \right) .  \tag{A10}
\end{equation}%
Finally, we replace this last expression in the right hand side of $\left(
A2\right) $ to arrive at the expression $\left( \ref{4.3}\right) .$ We put $z=w$
in $\left( \ref{4.3}\right) $ and we use the condition $\langle z;m,\varepsilon
\left\vert z;m,\varepsilon \right\rangle _{L^{2}\left( \mathbb{R}\right)
}=1. $ This allows us to obtain the expression $\left( \ref{4.4}\right) $ of the
normalization factor.
\\
\\
\begin{center}
\textbf{Appendix B}
\end{center}

\textbf{Proof}. For $x\in \mathbb{R},$ we can write successively 
\begin{align}
\mathcal{O}_{\varepsilon }\left[ \varphi \right] \left( x\right) &
=\sum_{n=0}^{+\infty }e^{-n\varepsilon }\langle \varphi \left\vert \varphi
_{n}\right\rangle \langle x|\varphi _{n}\rangle  \tag{B1} \\
& =\sum_{n=0}^{+\infty }e^{-n\varepsilon }\left( \int_{-\infty }^{+\infty
}\varphi \left( y\right) \overline{\langle y\left\vert \varphi
_{n}\right\rangle }dy\right) \langle x|\varphi _{n}\rangle  \tag{B2} \\
& =\int_{-\infty }^{+\infty }\left( \sum\limits_{n=0}^{+\infty
}e^{-n\varepsilon }\overline{\langle y\left\vert \varphi _{n}\right\rangle }%
\langle x|\varphi _{n}\rangle \right) \varphi \left( y\right) dy.  \tag{B3}
\end{align}%
We now look closely at the sum 
\begin{equation}
\mathcal{G}_{\varepsilon }\left( x,y\right) :=\sum\limits_{n=0}^{+\infty
}e^{-n\varepsilon }\overline{\langle y\left\vert \varphi _{n}\right\rangle }%
\langle x\left\vert \varphi _{n}\right\rangle =\sum\limits_{n=0}^{+\infty
}e^{-n\varepsilon }\varphi _{n}\left( x\right) \overline{\varphi _{n}\left(
y\right) }.  \tag{B4}
\end{equation}%
Recalling the expression $\left( \ref{3.7}\right) $ of the $\left\{ \varphi
_{n}\right\} $ then $\left( \ref{4.21}\right) $ reads 
\begin{equation}
\mathcal{G}_{\varepsilon }\left( x,y\right) =\frac{1}{\sqrt{\pi }}e^{-\frac{1%
}{2}\left( x^{2}+y^{2}\right) }\sum\limits_{n=0}^{+\infty }\left( \frac{1}{2}%
e^{-\varepsilon }\right) ^{n}\frac{1}{n!}H_{n}\left( x\right) H_{n}\left(
y\right) .  \tag{B5}
\end{equation}%
Eq.$\left( B5\right) $ can be rewritten as 
\begin{equation}
\mathcal{G}_{\varepsilon }\left( x,y\right) =e^{-\frac{1}{2}\left(
x^{2}+y^{2}\right) }K\left( e^{-\varepsilon };x,y\right)  \tag{B6}
\end{equation}%
where we have introduced the kernel function 
\begin{equation}
K\left( \tau ;x,y\right) :=\sum\limits_{j=0}^{+\infty }\tau ^{j}\frac{1}{j!}%
H_{j}\left( x\right) H_{j}\left( y\right) ;\quad 0<\tau <1.  \tag{B7}
\end{equation}%
The latter can be written in a closed form by applying the Mehler formula ( 
\cite{MOS}, p.252): 
\begin{equation}
K\left( \tau ;x,y\right) =\frac{\pi ^{-\frac{1}{2}}}{\sqrt{1-\tau ^{2}}}\exp
\left( \frac{2\tau }{1+\tau }xy-\frac{\tau ^{2}}{1-\tau ^{2}}\left(
x-y\right) ^{2}\right)  \tag{B8}
\end{equation}%
which also is the Poisson kernel for the Hermite polynomials expansion.
Taking this into account, Eq.$\left( B3\right) $ takes the form

\begin{equation}
\mathcal{O}_{\varepsilon }\left[ \varphi \right] \left( x\right) =e^{-\frac{1%
}{2}x^{2}}\int_{-\infty }^{+\infty }\varphi \left( y\right) e^{-\frac{1}{2}%
y^{2}}K\left( e^{-\varepsilon },x,y\right) dy.  \tag{B9}
\end{equation}%
We can also write the right hand side of $\left( B9\right) $ as 
\begin{equation}
\mathcal{O}_{\varepsilon }\left[ \varphi \right] \left( x\right) =e^{-\frac{1%
}{2}x^{2}}M_{\varepsilon }\left[ \varphi \right] \left( x\right) ,  \tag{B10}
\end{equation}%
where 
\begin{equation}
M_{\varepsilon }\left[ \varphi \right] \left( u\right) =\int_{-\infty
}^{+\infty }K\left( e^{-\varepsilon },x,y\right) \varphi \left( y\right) e^{-%
\frac{1}{2}y^{2}}dy.  \tag{B11}
\end{equation}%
This suggests us to introduce the function 
\begin{equation}
f\left( y\right) :=\varphi \left( y\right) e^{-\frac{1}{2}y^{2}},y\in 
\mathbb{R}.  \tag{B12}
\end{equation}%
which statisfies 
\begin{equation}
\left\Vert f\right\Vert _{L^{2}\left( \mathbb{\,R},e^{-y^{2}}dy\right)
}=\left\Vert \varphi \right\Vert _{L^{2}\left( \mathbb{R}\right) }. 
\tag{B13}
\end{equation}%
We now apply the result of B. Muckenhoopt $\left[ Mu\right] $ who considered
the Poisson integral of Hermite polynomials expansion and proved that for a
function $f\in L^{p}\left( \mathbb{R},e^{-y^{2}}dy\right) $ with $1\leq
p\leq +\infty $ the integral defined by 
\begin{equation}
A\left[ f\right] \left( \tau ,x\right) :=\int_{0}^{+\infty }K\left( \tau
,x,y\right) f\left( y\right) e^{-y^{2}}dy;\quad 0\leq \tau <1  \tag{B14}
\end{equation}%
with the kernel $K\left( \tau ,\bullet ,\bullet \right) $ defined as given
in $\left( B8\right) $ satisfies $\lim_{\tau \rightarrow 1^{-}}A\left[ f%
\right] \left( \tau ,y\right) =f\left( y\right) $ almost everywhere in $%
\left[ 0,+\infty \right[ ,1\leq p\leq \infty .$ We apply this result in the
case $p=2,$ $A\equiv M$ and $\tau =e^{-\varepsilon }$ to obtain that $%
M_{\varepsilon }\left[ \varphi \right] \left( x\right) \rightarrow e^{\frac{1%
}{2}x^{2}}\varphi \left( x\right) $, $a.e$. as $\varepsilon \rightarrow
0^{+} $ , which says that the limit $\mathcal{O}_{\varepsilon }\left[
\varphi \right] \left( x\right) =e^{-\frac{1}{2}x^{2}}M_{\varepsilon }\left[
\varphi \right] \left( x\right) \rightarrow \varphi \left( x\right) ,a.e$.
as $\varepsilon \rightarrow 0^{+}$ is valid for every $\varphi \in
L^{2}\left( \mathbb{R}\right) $. In other words, we get the limit 
\begin{equation}
\lim_{\varepsilon \rightarrow 0^{+}}\int\limits_{\mathbb{C}}\left\vert
z;m,\varepsilon \right\rangle \langle z;m,\varepsilon |d\mu _{m,\varepsilon
}\left( z\right) =\mathbf{1}_{L^{2}\left( \mathbb{R}\right) .}  \tag{B15}
\end{equation}%
in terms of Dirac's\textit{\ bra-ket} notation. This completes the proof.$%
\square \smallskip \smallskip $

\textbf{Acknowledgments.} I would like to thank L. D. Abreu for reading this
manuscript and for many useful remarks and comments.

\end{document}